\documentclass[11pt]{article}
\usepackage{moriond,epsfig}

\def\be{\begin{equation}}
\def\ee{\end{equation}}
\def\bea{\begin{eqnarray}}
\def\eea{\end{eqnarray}}

\begin{document}
\vspace*{4cm}
\title{The dark(er) side of inflation}

\author{ Gabriela Barenboim}

\address{ Departamento de Fisica Teorica-IFIC, University of Valencia, Spain}

\maketitle\abstracts{We present a new approach to quintessential inflation, in
which both dark energy and inflation are explained by
the evolution of a single scalar field. We
start from a simple scalar potential with both oscillatory
and exponential behavior.We employ the conventional reheating 
mechanism of new inflation, in which the
scalar decays to light fermions with a decay width that is
proportional to the scalar mass. Because our scalar mass is
proportional to the Hubble rate, this gives
adequate reheating at early times while shutting off at late
times to preserve quintessence and satisfy nucleosynthesis
constraints.
}

\section{Intoduction}

If we assume the correctness of the standard Friedmann equation
evolution, the existence of dark energy engenders two profound dilemmas.
The first is the cosmological constant problem, and the second is the
``why now?''\ problem. Quintessence models attempt to address the
second problem by introducing a very weakly coupled scalar field
whose potential and/or kinetic function have special properties.
One of the most successful approaches to 
quintessence \cite{Dodelson:2001fq,tracker}
is to combine tracking with an oscillating behavior in the quintessence
potential. In such models the equation of state parameter $w(z)$ has
a periodic component, leading to occasional periods of accelerated
expansion during epochs where $w(z)\simeq -1$.

It is natural in this context to ask whether the quintessence scalar
could replace the inflaton. The idea of quintessential
inflation has been examined by a number of 
authors \cite{Peebles:1998qn}-\cite{Sami:2004xk}.
The straightforward approach is to cobble together a
scalar potential which has both a flat, large vev portion
(for inflation) and a flat, small vev portion (for quintessence).
These features are connected by a steep step which corresponds
to a period of cosmic kination. As discussed 
in \cite{Dimopoulos:2001ix,Sami:2004xk}, such models suffer
from generic problems. First, they require significant
\textit{ad hoc} tuning to simultaneously produce the features
of inflation and quintessence. Second, they require a ``sterile''
inflaton, in order to avoid the decay of the putative
quintessence scalar at the end of
inflation. This in turn requires new mechanisms for reheating, such as
gravitational de Sitter phase particle production, leading
to difficulties in satisfying the constraints of CMB anisotropies
and of primordial nucleosynthesis.

\begin{figure}
   \hspace{4cm} \includegraphics[width=7cm]{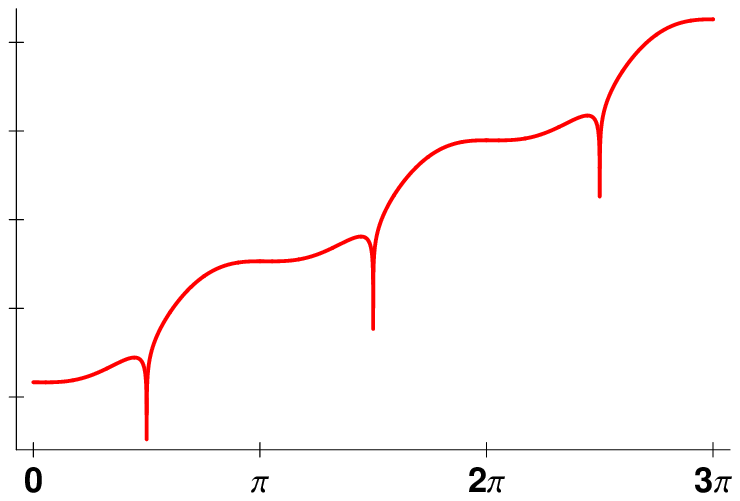}%{figure 1 } 
\newline

\vspace{1cm}

\hspace{4cm}\includegraphics[width=7cm]{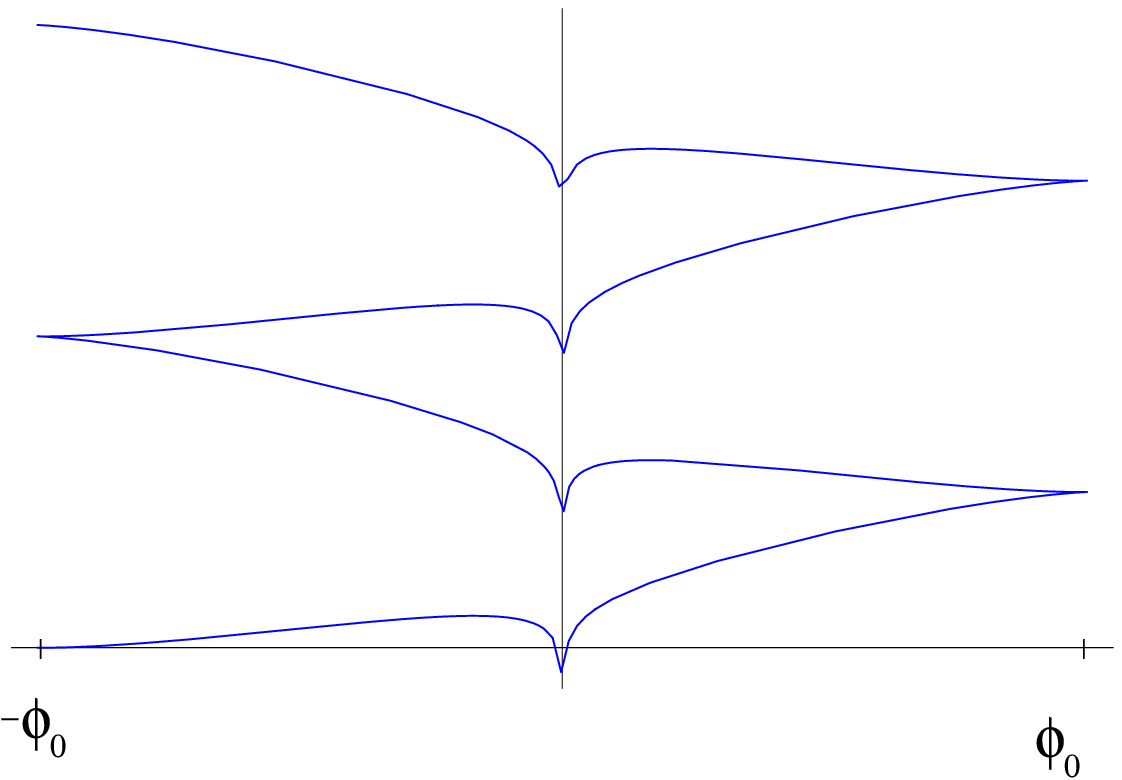}
\caption{The upper (lower) panel shows the 
scalar potential $V(\theta )$ ($V(\phi )$) for $b=1$, plotted
on a logarithmic scale.}
\label{fig:slinky}
\end{figure}  
 
Our approach to quintessential inflation 
\cite{Barenboim:2005np,Barenboim:2006rx}
is to take advantage
of the tracking and oscillatory potential features that work
so well in addressing the ``why now'' problem of quintessence alone  
\cite{Felder:1999pv}.
We will describe a model with a simple scalar lagrangian with exponential
and oscillatory features. 
The model uses the conventional reheating 
mechanism of new inflation \cite{Abbott:1982hn}, in which the
scalar decays to light fermions with a decay width that is
proportional to the scalar mass. We show that the scalar mass
is proportional to the Hubble rate. As a result, the model has
adequate reheating at early times while naturally shutting off at later
times.

We present a successful model with only three adjustable 
parameters. One parameter controls the period between 
inflationary epochs, a second controls the overall decay width,
and the third parametrizes our ignorance about the relative
fraction of matter versus radiation produced by reheating.
These three parameters are adjusted to produce sufficient
inflation along with the correct fractions
$\Omega_r/\Omega_{\Lambda}$, $\Omega_m/\Omega_{\Lambda}$ of
radiation, matter, and dark energy, as measured today.
Having thus fixed the model we find that we automatically
satisfy all constraints of primordial nucleosynthesis, CMB,
and large-scale structure.

\section{An oscillatory potential}
We start with a simple model with a single real
scalar quintessence field $\theta$. The action is
\begin{eqnarray}
\int d^4x \sqrt{-g} \left[
\frac{1}{2}f(\theta )g^{\mu\nu}\partial_{\mu}\theta
\partial_{\nu}\theta - V(\theta ) \right] \; ,
\label{eqn:piaction}
\end{eqnarray}  
where the kinetic function $f(\theta )$ and potential
$V(\theta )$ are given by:
\begin{eqnarray}\label{eqn:ourkin}
f(\theta ) &=& {3M_{\rm P}^2\over\pi b^2}\, {\rm sin}^2\,\theta\; ;\\
V(\theta ) &=& \rho_0\, {\rm cos}^2\,\theta\, {\rm exp}\left[{3\over b}\left(
2\theta - {\rm sin}\,2\theta \right) \right] \; ,
\label{eqn:ourpot} 
\end{eqnarray}
where $M_{\rm P}$ is the Planck mass: 
1.22 $\times 10^{19}$ GeV; $\rho_0$ is the dark energy
density observed today: $\simeq$ ($10^{-4}$ eV)$^4$; $b$ is a
dimensionless parameter which controls the periodic behavior.
A canonical kinetic term can be restored via a field redefinition
$\theta (x) \to \phi (x)$, where 
\begin{eqnarray}
\phi (x) = \phi_0\, {\rm cos}\,\theta \; ,
\label{eqn:canonical}
\end{eqnarray}
with $\phi_0\equiv \sqrt{3M_{\rm P}^2/\pi b^2}$.

In the approximation where we ignore the energy density
of radiation and matter, 
and where the only friction is from the metric expansion,
the evolution of the model can be solved
analytically. Energy conservation requires:
\begin{eqnarray}
\dot{\rho}_{\theta} = -3H(1+w)\rho_{\theta} \; ,
\end{eqnarray}
where $H$ is the Hubble rate, $w$ is the equation of state
parameter, and $\rho_{\theta}$ is the dark energy density.
The solution to this equation as a function of 
the scale factor $a(t)$ is:
\begin{eqnarray}
\rho_{\theta}(a) = \rho_0 \, {\rm exp}\left[
-3\int_1^a {da\over a}(1+w(a)) \right] \; ,
\label{eqn:rhorel}
\end{eqnarray}
where $\rho_0$ is the dark energy density at $a=1$ (today).

We also know that
\begin{eqnarray}
V(\theta) = \frac{1}{2}(1-w)\rho_{\theta} \; .
\label{eqn:vrhorel}
\end{eqnarray}
Making the Ansatz 
\begin{eqnarray}
w(a) = -{\rm cos}\,2\theta (a) \; ,
\end{eqnarray}
one immediately gets a solution to the 
(flat) Friedmann
equation combined with the
relations (\ref{eqn:rhorel}-\ref{eqn:vrhorel}):
\begin{eqnarray}\label{eqn:ourtheta}
\theta (a) &=& -{b\over 2}\,{\rm ln}\,a \; ;\\
w(a) &=& -{\rm cos}\,[b\,{\rm ln}\,a] \; .
\label{eqn:ourw}
\end{eqnarray}

The expectation value
of the quintessence field $\theta$ evolves logarithmically with scale
factor from a positive initial value to zero today. 
Accelerated expansion
corresponds to epochs (such as today)
where $\theta$ is evolving through one of the
flat ``steps'' of the potential. From (\ref{eqn:ourkin}) we
see that the kinetic function is simultaneously suppressed
in this epochs, slowing the roll of the scalar field evolution.

The equation of state parameter $w(a)$ has the 
same periodic form assumed in  recent phenomenological 
analyses \cite{Barenboim:2004kz,Barenboim:2005wx,Feng:2004ff}. The analysis 
in \cite{Barenboim:2004kz}
showed that, for $b=1$, $w(a)$
is consistent with all current data from observations of
the CMB,
Type IA supernovae, and large scale structure.
It follows \textit{a fortiori} that our model with any choice
of $b$ less than one also agrees with this data.

To complete the model, we will assume that the quintessence
field $\phi$ has a weak perturbative coupling to light fermions.
This is the standard reheating mechanism of new 
inflation \cite{Abbott:1982hn}.
To avoid the strong constraints on long-range forces mediated 
by quintessence scalars \cite{Carroll:1998zi},
it is simplest to imagine that our
scalar only has a direct coupling to a sterile neutrino.
This is sufficient to hide the quintessence force from
Standard Model nonsinglet particles \cite{Fardon:2003eh},
while still allowing the 
generation of a radiative thermal bath of Standard Model
particles from quintessence decay.

With this assumption for reheating the evolution equations
for quintessence, radiation, and matter become:
\begin{eqnarray}
\dot{\rho}_{\theta} &=& -3H (1+w) \rho_{\theta} 
-k_0m_{\phi}(1+w)\rho_{\theta}
\; ;\nonumber\\
\dot{\rho}_r &=& -4H \rho_r 
+(1\hspace*{-2pt}-\hspace*{-2pt}f_m)k_0m_{\phi}(1+w)\rho_{\theta}
\; ; \\
\dot{\rho}_m &=& -3H \rho_m +f_mk_0m_{\phi}(1+w)\rho_{\theta}
\; ;\nonumber
\label{eqn:coupled}
\end{eqnarray}
where $k_0$ and $f_m$ are small dimensionless constants.
As long as $\theta$ is not near
a multiple of $\pi/2$, it is a reasonable approximation to make
the replacement $ k_0m_{\phi} \to kH \; $
where $k$ is another small dimensionless parameter.
This replacement decouples the $\theta$ evolution equation
from the Friedmann equation, giving an immediate analytic
solution:
\begin{eqnarray}
\rho_{\theta}(a) = \rho_0 \, 
{\rm exp}\left[{1\over b}(3+k)\left(
2\theta - {\rm sin}\,2\theta \right) \right]
\; ,
\label{eqn:rhonew}
\end{eqnarray}
We have used this approximation in the solutions quoted
below.

\begin{figure}
\begin{center}
    \includegraphics[width=8cm]{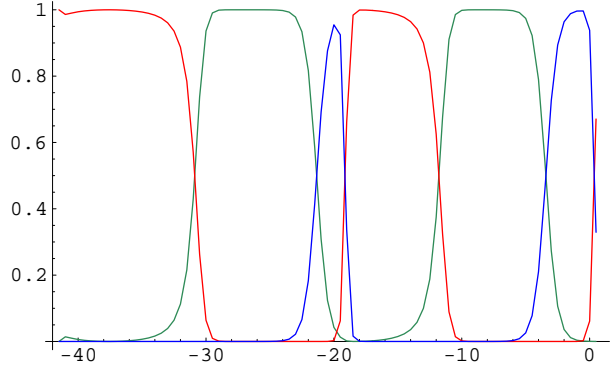}%{figure 2}    
\vspace*{-2pt}
\end{center}
\caption{History of a slinky
universe. Shown are the relative energy density
fractions in radiation (green), matter (blue), and 
dark energy (red), as a function of the logarithm
of the scale factor $a(t)$.}
\label{fig:history}
\end{figure}  
\begin{table}[h]
\centering
\begin{tabular}{|c|c|c|c|c|c|c|}
\hline\hline
log$_{10}$ $a $ & -42 & -40 & -38 & -36 & -34 & -32 \\ [0.5 ex]
\hline
$w(a)$ & -0.32 & -0.83 & -1 & -0.75 & -0.19 & 0.45 \\ \hline
$T$ & 0 & 2 10$^{17}$ & 3 10$^{16}$ & 10$^{17}$ & 2 10$^{16}$ & 8 10$^{14}$ \\ \hline
$\Omega_\Lambda$ & 1 & 0.997 & 0.99998 & 0.996 &
0.976 & 0.79 \\ \hline
$\Omega_r$ & 0 & 0.003 & 2 10$^{-5}$ & 0.004 & 0.024 & 0.21 \\ \hline
$\Omega_m$ & 0 & 5 10$^{-14}$ & 3 10$^{-16}$ & 5 10$^{-14}$
& 4 10$^{-13}$ & 2 10$^{-11}$   \\\hline
log$_{10}$ $a $ & -30 & -28 & -26 & -24 & -22 & -20 \\\hline
$w(a)$ & 0.90 & 0.98 & 0.64 & 0.04 & -0.58 & -0.96 \\ \hline
$T$ & 9 10$^{12}$ & 9 10$^{10}$ & 9 10$^8$ & 9 10$^{6}$ & 9 10$^{4}$ & 900  \\ \hline
$\Omega_\Lambda$ & 0.01 & 8 10$^{-7}$ & 5 10$^{-10}$ & 
2 10$^{-10}$ & 5 10$^{-7}$ & 0.07 \\\hline
$\Omega_r$ & 0.99 & 0.999998 & 0.99993 & 0.993 & 0.59 & 0.01 \\\hline
$\Omega_m$ & 7 10$^{-9}$ & 7 10$^{-7}$ & 7 10$^{-5}$
& 0.007 & 0.41 & 0.92   \\\hline
log$_{10}$ $a $ & -18 & -16 & -14 & -12 & -10 & -8 \\\hline
$w(a)$ & -0.94 & -0.52 & 0.11 & 0.69 & 0.99 & 0.87 \\ \hline 
$T$ & 220 & 170 & 17 & 0.3 & 3 10$^{-3}$ & 3 10$^{-5}$  \\ \hline
$\Omega_\Lambda$ & 0.9992 & 0.99 & 0.94 & 0.33 & 0.0002
& 2 10$^{-8}$ \\\hline
$\Omega_r$ & 0.0008 & 0.01 & 0.06 & 0.67 & 0.9998 & 0.9999  \\\hline
$\Omega_m$ & 2 10$^{-5}$ & 5 10$^{-10}$ & 3 10$^{-11}$
& 5 10$^{-9}$ & 8 10$^{-7}$ &  10$^{-4}$ \\\hline
log$_{10}$ $a $ & -6 & -4 & -2 & 0 &  &  \\\hline
$w(a)$ & 0.39 & -0.25 & -0.79 & -1.0 &  &  \\ \hline
$T$ & 3 10$^{-7}$ & 3 10$^{-9}$ & 3 10$^{-11}$ & 3 10$^{-13}$ &  &   \\ \hline
$\Omega_\Lambda$ & 10$^{-10}$ & 2 10$^{-9}$ & 
7 10$^{-6}$ & 0.71 &  &   \\\hline
$\Omega_r$ & 0.992 & 0.56 & 0.01 & 0.00005 &  &  \\\hline
$\Omega_m$ & 0.008 & 0.44 & 0.99 & 0.29 &  &  \\[1ex]
\hline\hline
\end{tabular}
\caption{Relative density fractions of dark energy, radiation,
and matter, as a function of the scale factor. Also shown are the
temperature $T$ in GeV, and the equation of state parameter $w(a)$.}
\end{table}

\section{Results}

Figure \ref{fig:history} shows the results obtained
from our model with $b=1/7$, $k=0.06$, and
$f_m = 10^{-11}$. Shown are the relative energy density
fractions in dark energy, radiation, and matter, as a function
of ${\rm log}\,a$. We have chosen to integrate the
evolution equations starting from $a= 10^{-42}$, which in
our model corresponds to a temperature of slightly
less than  $10^{16}$ GeV, and an initial comoving Hubble
radius of about 100 Planck lengths. 
For simplicity we
have also chosen the value of $w(a)$ now to be exactly $-1$. 
Neither of these choices corresponds to a necessary tuning.

Our three adjustable parameters have been chosen such
that the values of  
$\Omega_r/\Omega_{\Lambda}$, $\Omega_m/\Omega_{\Lambda}$
come out to their measured values at $a=1$, and such that
we have sufficient inflation. The latter is checked by
computing the ratio of the fully inflated size of the
intitial comoving Hubble radius to the current comoving
Hubble radius. This ratio is about 3 in our model, indicating
that the total amount of inflation is indeed enough to solve
the horizon problem. The flatness problem is solved because
the total $\Omega_r+\Omega_m+\Omega_{\Lambda} =1$ within
errors.

\begin{figure}
\centerline{\epsfxsize 5.5 truein \epsfbox {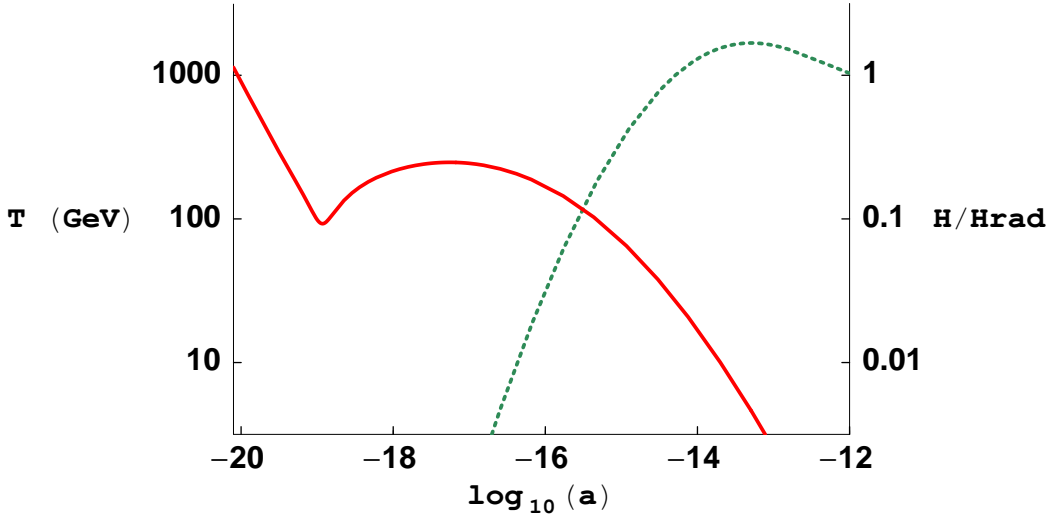}}
\caption{\label{fig:modeltwotemp} The temperature history of
Model 2 (red/solid) near the electroweak phase transition. Shown in
green/dashed is the Hubble parameter $H$ of Model 2 normalized to
the expansion rate $H_{rad}$ for pure radiation.
\hbox to170pt{}}
\end{figure}

From Figure (\ref{fig:history}) we see that we are currently
beginning the third epoch of accelerated expansion. The first
epoch of inflation accumulated about 18 $e$-foldings. Quantum
fluctuations during this epoch produced the spatial inhomogeneities 
responsible for large scale structure and CMB anisotropies
observed today. 
Constraints on the physics responsible for the primodial power spectrum 
of these density fluctuations can be set with WMAP and 2dF data,
under the assumption that the Hubble rate is dominated by the
contribution from $\rho_{\phi}$
during the observable part of inflation \cite{Schwarz:2004tz}.
As can be seen from the figure a 
second period of accelerated expansion began just before
the electroweak phase transition, and ended well before BBN.
The temperature history near the EWPT is shown in Figure \ref{fig:modeltwotemp}.

Also shown is the Hubble parameter $H$ of the model normalized to
the expansion rate $H_{rad}$ for pure radiation. $H_{rad}$ corresponds
to what is assumed in the standard paradigm. Notice that for temperatures
of a few GeV the expansion rate is actually somewhat larger than normal,
but at higher temperatures it is much less than normal.

Such a nonstandard thermal history will impact on electroweak baryogenesis.
For a Higgs sector such that the EWPT is first order, the change in the net
baryon asymmetry is proportional to -log$(H/H_{rad})$, where $H$
is the expansion rate during the phase transition, and $H_{rad}$ is
the corresponding expansion rate for pure radiation \cite{Joyce:1997fc}.
If the Higgs sector is such that the EWPT is second order, the
baryon asymmetry is proportional to the expansion rate \cite{Joyce:1997fc}.
Clearly one should revaluate the popular scenarios
for electroweak baryogenesis in this light.

Such a model will have major implications for 
predictions of the relic abundance of dark matter particles with Terascale
masses. The dominant production mechanism for such particles may
be scalar decays, as suggested by Figure \ref{fig:history}. Even if the
dark matter particles are thermal relics, their abundance now will be
affected by the nonstandard expansion rates at 
earlier times.

To compare with contraints from primordial nucleosynthesis
\cite{Bean:2002sm},
we note that the second epoch of dark energy domination
ended well before $a \simeq 10^{-10}$, the time at which
nucleosynthesis occured. Indeed dark energy reheating effects
are completely negligible from $a \simeq 10^{-10}$ until today.

Figure (\ref{fig:history}) also shows one (very brief)
prior epoch of matter domination. This only occurs if there
is some suitably heavy and long-lived matter around to
go out of thermal equilibrium when $a \sim 10^{-22}$,
\textit{e.g.} a superheavy neutrino.

It is difficult to extract a precise prediction for
the spectral indices of this model,
since we are never strictly in the slow roll regime.
We will be content here with a rough estimate.
This is obtained starting from a canonical field
redefinition:
\begin{eqnarray}
\phi (x) = 2\sqrt{3}\; \frac{k}{b}\,{\rm cos}\,\theta \; .
\end{eqnarray}
In terms of the canonical scalar $\phi$, the potential
\ref{eqn:ourpot} can be written:
\begin{eqnarray}
V(\phi ) &\propto&  \phi^2 \;  H^2 (\phi ) \; ,
\end{eqnarray}
where $H(\phi )$ is the Hubble rate we would get
ignoring radiation and matter.
During inflation, $H$ is approximately constant,
but this is not an especially good approximation since
we are not in a slow roll regime. This is similar to
the oscillatory models of quintessential inflation
discussed in \cite{Nunes:2002wz}. Taking the potential
in the inflationary phase to be approximated by
$V \propto \phi^2$, we can estimate the
scalar spectral index $n_s$:
\begin{eqnarray}
n_s  \simeq 1 -\frac{2}{N}
\end{eqnarray}
with $N$ the number of e-folds between the Hubble radius exit and
the end of the second inflationary period.
For our model $ N = 29$, giving $n_s = .93$, in good agreement with
recent observations.

It is fair to say that, compared to the simple model
presented above, the standard new inflation scenario looks
rather extreme. In the evolution history
portrayed in Figure \ref{fig:history}, the interplay between
inflationary expansion and reheating is much milder. In fact,
apart from a few very brief periods, reheating effects in
our model do not actually increase the temperature of the
thermal radiation bath. Instead, the temperature is almost
always decreasing, but it decreases more slowly than in
the standard $\Lambda$CDM evolution. 

Our particular realization of Slinky inflation must
be regarded with some care,
since $\phi_0 > M_{\rm P}$ (as in all models of chaotic
inflation) and
we have not invoked any consistent Planck scale framework
such as string theory. However the model itself
is surprisingly simple, and the physical picture 
which emerges from it has some compelling features.
These are worthy of further investigation. Also,
since the target space parametrized
by our scalar is $\mathbf{S^1}$, it would be
interesting to extend this scenario to a class of nonlinear
sigma models with other compact target spaces.

\section*{Acknowledgments}
It is a pleasure to thank Joe Lykken in collaboration with whom,
the work reported here has been done. 
GB is grateful for support from the Spanish MEC and FEDER under 
Contract FPA 2005/1678,
and to the Generalitat Valenciana under Contract GV05/267.

\end{document}